# Analysis of Transcranial Focused Ultrasound Beam Profile Sensitivity for Neuromodulation of the Human Brain

Jerel K. Mueller[1], Wynn Legon[2], and William J. Tyler*[3]

[1] School of Biomedical Engineering and Sciences, Virginia Tech, Blacksburg, VA
[2] Dept. of Physical Medicine and Rehabilitation, University of Minnesota, Minneapolis, MN
[3] School of Biological and Health Systems Engineering, Arizona State University, Tempe, AZ

*Correspondence: wtyler@asu.edu

**Abstract**

*Objective.* While ultrasound is largely established for use in diagnostic imaging and heating therapies, its application for neuromodulation is relatively new and not well understood. The objective of the present study was to investigate issues related to interactions between focused acoustic beams and brain tissues to better understand possible limitations of transcranial ultrasound for neuromodulation. *Approach.* A computational model of transcranial focused ultrasound was constructed and validated against bench top experimental data. The models were then incrementally extended to address and investigate a number of issues related to the use of ultrasound for neuromodulation. These included the effect of variations in skull geometry and gyral anatomy, as well as the effect of transmission across multiple tissue and media layers, such as scalp, skull, CSF, and gray/white matter on ultrasound insertion behavior. In addition, a sensitivity analysis was run to characterize the influence of acoustic properties of intracranial tissues. Finally, the heating associated with ultrasonic stimulation waveforms designed for neuromodulation was modeled. *Main results.* Depending on factors such as acoustic frequency, the insertion behavior of a transcranial focused ultrasound beam is only subtly influenced by the geometry and acoustic properties of the underlying tissues. *Significance.* These issues are critical for the refinement of device design and the overall advancement of ultrasound methods for noninvasive neuromodulation.

1. Introduction

Transcranial-focused ultrasound (tFUS) is an emerging technology for non-surgical stimulation of the human brain. tFUS offers a superior millimeter resolution compared to existing technologies like transcranial magnetic stimulation (TMS) which influences areas of the cortex spanning several centimeters [1, 2]. Recently, it has been demonstrated that tFUS directed over the somatosensory cortex in humans affects EEG amplitude, power, phase, and tactile behavior [3, 4]. The intracranial manifestation of mechanical and thermal effects by tFUS depends on the insertion behavior of ultrasound across the various layers of tissue. Additionally, it is still not clear how the neuronal response couples to the exertions of ultrasound on neural tissue. An understanding of both the insertion behavior of ultrasound across the tissue layers in the context of neuromodulation and the coupled neuronal response is key to the continued advancement of ultrasound stimulation methods. The objective of the present study was to investigate the insertion behavior of tFUS for neuromodulation across the human head and quantify its sensitivity to tissue domains, their parameters, and their geometry.

Focused ultrasound has previously been investigated for such applications as brain tumor ablation, blood-brain barrier opening, and thrombolysis [5]. In these applications, it is advantageous to deliver the desired level of ultrasound energy through an intact human skull to the prescribed locations, especially for deeper subcortical regions. The intact skull though represents the primary barrier to ultrasound. The high attenuation, diffusion, and refraction of ultrasound waves in cranial bone compared to the neighboring tissues results in a significant loss of energy and distortion of the transmitted ultrasound beam, and is the primary barrier to high resolution transcranial ultrasound imaging [6]. To an extent, adaptive focusing techniques are able to account for the defocusing effect of the skull [7], and is critical to the application of high intensity focused ultrasound. In the context of neuromodulation though, despite bone absorbing ultrasound almost 90 times more efficiently than soft tissue [8], the skull does not pose such a dire obstacle to the transmission of sufficient energy for low intensity focused ultrasound applications. In addition to the effect of tissue properties on ultrasound, it is also important to demonstrate that ultrasound for neuromodulation does not heat the tissue. At low intensities over short exposure times, ultrasound does not generate appreciable tissue heating, and the mechanical effects of ultrasound used in neuromodulatory capacities has not been reported to cause tissue damage [9-11]. Thus, as the safety of ultrasound has been extensively investigated, and the insertion behavior of ultrasound characterized in the context of various other applications, there is a need to explore the insertion behavior of tFUS and the heating characteristics for the purposes of neuromodulation beyond the barrier of the skull. We developed a computational model of tFUS for neuromodulation and used this model to explore the insertion behavior of the ultrasound beam in the intracranial space. We evaluated several paradigms to explore the sensitivity of focused ultrasound to tissue layers, their acoustic properties, and their geometry.

## 2. Methods

We developed computational models of the human skull and superficial cortical layers, including CSF, white matter, and gray matter, to evaluate the insertion behavior of tFUS across the skull and the resultant intracranial maps of intensity and heating. The finite element method models were constructed in COMSOL Multiphysics v4.3 (COMSOL, Burlington, MA) to calculate pressure, intensity, and heat generation. By these methods, we were able to investigate the subtle influence that various aspects of human biology impart on the behavior of tFUS.

The initial computational model recreated quantitative acoustic field mapping of focused ultrasound transmitted through a hydrated fragment of human cranium, which has been detailed previously [3]. Briefly, a calibrated hydrophone mounted on a motorized stage was used to measure the acoustic intensity profile from the ultrasound transducer coupled to a skull fragment in a 58 L acrylic water tank at a 400 µm spatial resolution. The ultrasound transducer is a custom designed single-element focused transducer (Blatek, Inc., State College, PA) having a center frequency of 0.5 MHz, a diameter of 30 mm, and a focal length of 30 mm. The transcranial ultrasonic neuromodulation waveform used has been previously described [12, 13], and has an acoustic frequency of 0.5 MHz, a pulse duration of 360 µs, and consists of 500 pulses delivered at a pulse repetition frequency of 1.0 kHz, resulting in a stimulus duration of 0.5 sec. As reported in Legon et al. 2014, we observed that transcranial transmission using this setup results in a spatial-peak pulse-average intensity ($I_{SPPA}$) of 5.90 W/cm$^2$ [3].

To recreate this experiment in the computational environment, a two-dimensional geometry with axial symmetry was created as shown in figure 1a. The left most edge was specified as the axis of rotation, and the bottom most circular edge was specified as the single element of the focused transducer which serves as the ultrasound source. The transducer element is shaped with a focal length of 30 mm and an aperture diameter of 30 mm, simulating the transducer used in bench top experiments within the water tank, and is similarly driven at a frequency of 0.5 MHz. The normal displacement of the transducer element was specified as 13 nm, based on calculations involving the piezoelectric constant of the transducer element materials and the ultrasonic neuromodulation waveform. Above the transducer element is a plane of skull 5 mm thick, and the space between the transducer and skull was specified as water. Beyond the skull layer the space was specified as water again, to recreate the conditions of the bench top measurements. The material properties specified in each domain are detailed in table 1, and geometries detailed in table 2. The sound velocity, density, thermal conductivity, and heat capacity of the water domain is derived from the default material properties for water in COMSOL.

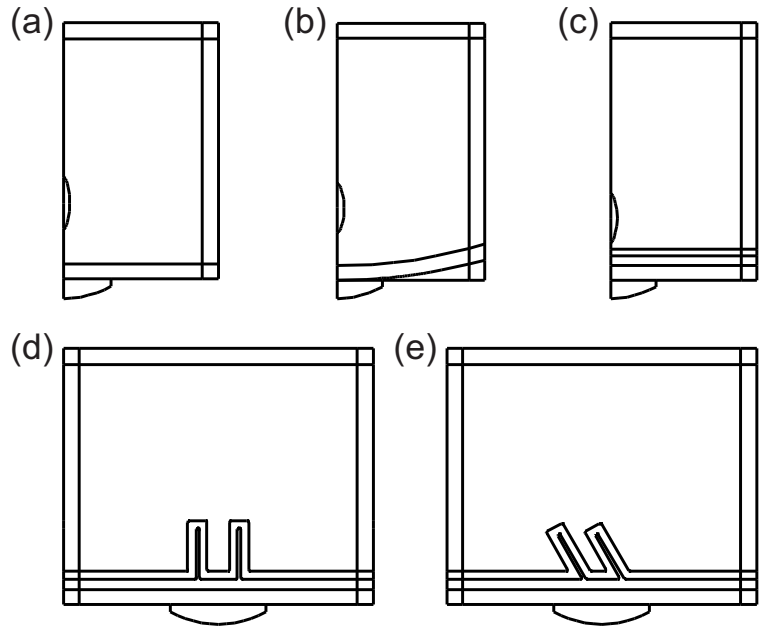

**Figure 1. Geometries of computational models of transcranial focused ultrasound.** (a) Water tank model. (b) Curved skull model. (c) Layered brain model. Separate layers account for the skull, CSF, gray matter, and white matter. (d) Perpendicular sulci model. (e) Slanted sulci model. Sulci are rotated 30° from the perpendicular.

**Table 1.** Material parameters.

| Material | Speed of sound (m/sec) | Density (kg/m$^3$) | Attenuation Coefficient (Np/m) | Thermal Conductivity (W/K/m) | Heat Capacity (J/kg/K) |
|---|---|---|---|---|---|
| Water | 1,483 | 999.5 | 0.02 | 0.595 | 4,186 |
| Skull | 2,300 [14] | 1,912 [14] | 21.5 [15] | 0.43 [16] | 1,440 [16] |
| CSF | Water | Water | Water | Water | Water |
| Brain – GM | 1,550 [17] | 1,030 [17] | 0.92 [16] | 0.528 [16] | 3,640 [16] |
| Brain - WM | 1,550 [17] | 1,030 [17] | 0.92 [16] | 0.528 [16] | 3,640 [16] |

Gray Matter (GM); White Matter (WM); Cerebrospinal Fluid (CSF)

Table 2. Model geometry.

| Object | Dimensions |
|---|---|
| Axial model span | 50 mm radial / 80 mm height |
| Gyral model span | 100 mm width / 80 mm height |
| Skull | 5 mm thick [18] |
| CSF | 3.1 mm thick |
| Gray matter | 2.5 mm thick [19] |
| Precentral gyrus | 12 mm wide [18] |
| Central sulcus | 1 mm wide / 16 mm deep [18] |
| Precentral sulcus | 1 mm wide / 16 mm deep [18] |

To more finely resolve the pressure gradients in the focal area of the transcranial domain, the mesh size was specified as 1/6 of the wavelength within an elliptical region enclosing the focal point. A coarser mesh of 1/4 of the wavelength was specified in all other regions. Additionally, the model is bounded along the top and sides by cylindrical perfectly matched layers to absorb the outgoing ultrasound waves and prevent their reflection back into the modeling domain. The model solved for the stationary acoustic field to determine the acoustic intensity distribution in the materials with the further assumptions that acoustic wave propagation is linear and that the amplitude of shear waves are nominal compared to those of the primary, compressive waves. Shear waves are greatly attenuated by tissue compared to the longitudinal waves in an ultrasound beam [20, 21].

The acoustic intensity magnitude was then used to calculate the heat source for thermal simulations. Material properties were assumed to not change with temperature and that cooling due to blood perfusion was negligible in model layers of biological tissue. For the computational models recreating bench top experiments in the water tank, all domains were assigned an initial temperature of 294 K, corresponding to room temperature. For computational models investigating the stimulation and heating of biological tissue, all tissue domains were assigned an initial temperature of 310 K, corresponding to body temperature. The time course of application of the heat source was specified to mimic the transcranial ultrasonic neuromodulation waveform. Heating was applied in 360 µs durations repeated at 1.0 kHz for 0.5 sec. No heating was applied for an additional 0.5 sec to observe cooling of the tissue, bringing total simulation time to 1 sec. The max time step of thermal simulations was set at 1/7 of the pulse duration, to resolve the heating and cooling between each application of an ultrasound pulse.

Following comparison of the experimental acoustic field mapping to the representative computational model, the model was then expanded to investigate how features such as geometry and material properties influence the insertion behavior of transcranially-focused ultrasound in humans. First the transcranial domain was given material properties of brain tissue, as specified for gray and white matter in table 1. This offers a simplified model of the effects of transcranial-focused ultrasound in humans and a baseline for comparisons to later models. Next the skull layer was curved (figure 1b), given a radius of 17 cm, to deviate from the idealized straight plane of the previous models, and decrease the effective mechanical coupling between the transducer and skull.

Tissue layers for cerebrospinal fluid (CSF), gray matter, and white matter were then added above the skull layer (figure 1c) with the thicknesses and material properties stated in table 1 and table 2. The thickness of layers was based on computational models of electrical epidural motor cortex stimulation [19] and is specified in table 2. The thickness of the CSF layer was derived from the sum of the thicknesses of the dura mater and CSF layers from previous models of the precentral gyrus [18], as we were unable to find the relevant acoustic parameters for the dura mater in literature for our models. Additionally, the CSF was assumed to have the material properties of water, due to the lack of literature characterizing the parameters of interest for our models. Sensitivity analyses were then run with the layered tissue model to further explore the influence of these additional layers. Models scaling the attenuation coefficient of white and gray matter by 0.1, 0.2, 0.5, 1.0, 1.4, 2.0, 5.0, and 10.0 were solved to inspect its influence on the ultrasound beam profile and tissue heating. The scaling factor 1.4 was included as the attenuation coefficient of white matter has been reported to be 1.4 times that of gray matter [22]. Additionally, the thickness of the CSF layer was scaled by 0.1, 0.2, 0.5, 1.0, 2.0, and 5.0 to inspect the influence of the material's presence as CSF volume is known to vary, such as due to age related loss of cortical volume [23]. Furthermore, sensitivity analyses scaling the density and speed of sound properties of the gray and white tissue layers were run, as the mechanical properties of cortical tissue have been shown to vary with age and disease [24-26].

To gain insight on the influence of gyral geometry on the behavior of tFUS, we constructed two-dimensional models of the precentral gyrus, including two adjacent sulci and two neighboring gyri based on computational models of electrical epidural motor cortex stimulation [19]. In addition to models with the sulci oriented perpendicularly to the skull (figures 1d), other models with sulci slanted thirty degrees were constructed (figure 1e). Maximum element size within the gyral anatomy model domains was restricted to 1/6 the wavelength. Additionally, bounding perfectly matched layer domains surrounded the modeling area on all sides as ultrasound intensity profiles were solve for 0, 3, 6, 9, and 12 mm transducer offsets from the center line of the model. This modeling geometry assumes the plane of the model is the plane of symmetry, representing the middle plane of an infinite slab. Attempts to extend the model to three dimensions proved exceedingly computationally intensive and thus the model was kept two-dimensional. While this planar, infinite slab, geometry is not representative of the true geometry of the focused ultrasound transducer included on the bottom of the models, the resultant pressure field is qualitatively representative of the beam profile at the center plane of symmetry for the true geometry of the ultrasound transducer, and thus intensity profiles examined as normalized quantitates. Furthermore, to allow fair comparison to a model without gyral anatomy, another computational model similar to that of the layered cortex in figure 1c was constructed but without a plane of symmetry, thus resembling the geometries of figures 1d and 1e, but without the sulci.

     To quantify the area of stimulation by the focused ultrasound beam, the geometry of the root mean squared intensity ($I_{RMS}$) solved for from the FEM model was characterized for intensities greater than the half maximum. This thresholding resulted in an elliptical profile that served as the proxy for stimulated neural tissue in the computational model from which the area was calculated, excluding any area that was not contained in the gray or white matter (e.g. the CSF and skull). Additionally, the centroidal principal axes of the area moment of inertia of the thresholded intensity profile was calculated to determine their principal angles to characterize any deformation of the ultrasound beam. The length of the centroidal principal axes bounded by the thresholded profile was also determined to help characterize the geometry of the proxy for stimulated neural tissue.

## 3. Results

### 3.1. Computational model of acoustic water tank measurements

Calculations of intensity from a computational model were compared to experimental measurements. A close up of experimental measurements in the region of focus and minimally offset from the intracranial surface of the skull are shown in figure 2a and the intensities for a similar area in the computational model shown in figure 2b. Characterization of the half maximum intensity profile for both the computational and experimental models is shown in figure 2c and 2d. To allow comparisons between the experimental and computational fields, calculations of intensity within the skull layer were removed, as in figure 2b, and then the experimental and computational models aligned according to the location of maximal intensity. The absolute difference and relative error between the two data sets were then calculated, and is shown in figure 3. The greatest absolute differences are in a small region immediately adjacent to the skull, while all other regions, particularly at the focal region of the ultrasound transducer, are notably low. These absolute differences are further reflected in the calculations of relative error, figure 3b, where the errors are minimal solely in the focal region. The increase in relative error outside the focal region is attributable to intensity values approaching zero in the denominator of relative error calculations once outside of the focal region. Beyond the few differences in intensity profiles between experimental and computational models likely due to differences between ideal simulations and non-ideal observations, the qualitative similarity, and particularly the good quantitative agreement within the region of focus, between the computational and experimental profiles is reassuring of the model.

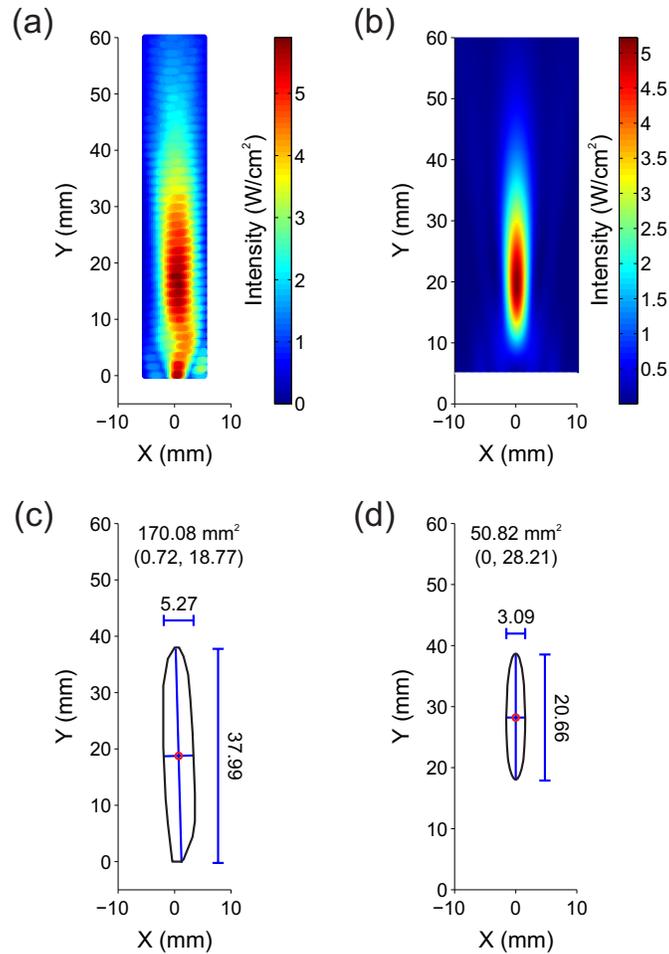

**Figure 2. Experimental and computational models of tFUS into a water domain.** (a) Experimental $I_{SPPA}$ measurements of transcranial focused ultrasound in a water tank. Skull border begins at a Y level of 0 mm. (b) Calculations of $I_{RMS}$ in a computational model recreating the experimental setup. Skull border begins at a Y level of 5 mm. (c) Half maximum intensity profile and characterization of the experimental measurements. Included on the plot is the area of the elliptical contour, the centroid of the contour, and the length of the centroidal axes bounded by the contour. (d) Half maximum intensity profile and characterization of the computational model.

Mueller et al., arXiv 2015 | **Page 10**

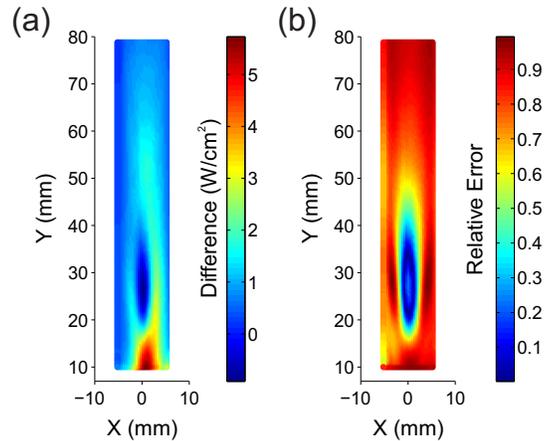

**Figure 3. Comparison of the intensity field between models.** (a) Calculation of the difference between the intensities of the computational and experimental models. Note that at the region of focus the differences are lowest, while differences are greatest at an off center location close to the skull. (b) Relative error between the intensities of the computational and experimental models. Note that within the region of focus that the relative error is low, while outside of the region of focus the error increases. This is due to the denominator of the calculation being composed of low intensity values that outside the region of focus are less than one.

The computational model also allows visualization of the profile of heat generation by tFUS, which is not as readily observed in experimental preparations, and is shown in figure 4. The heat generated in the skull is several orders of magnitudes greater than that generated in the water domain, and follows the profiles of intensity from tFUS. The time course of temperature change is shown in figure 4c. During US stimulation the temperature steadily rises with a rate highly dependent on the spatial location relative to the focus of the ultrasound transducer and the properties of the material exposed to ultrasound, with skull tissue heating up considerably more than the transcranial water domain.

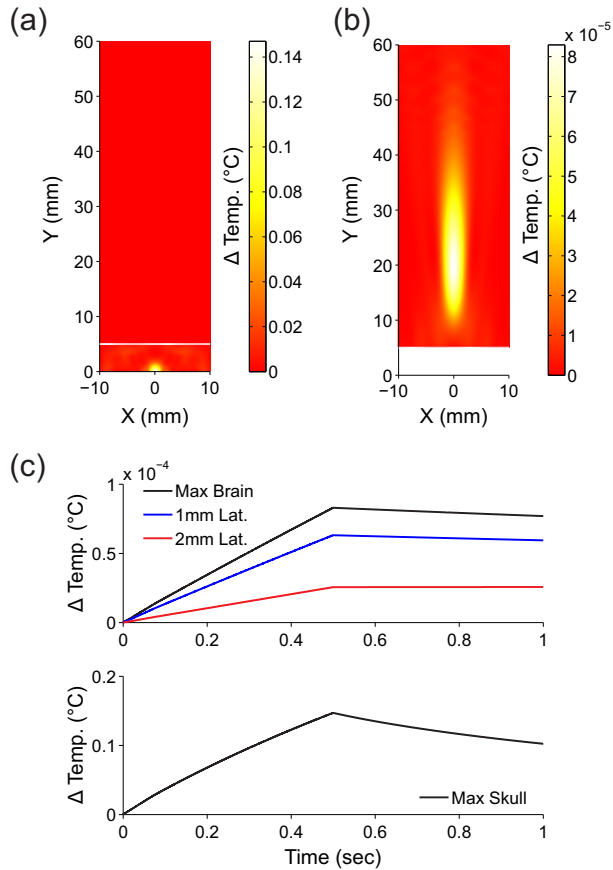

**Figure 4. Calculations of heat generation in the computational model of tFUS into a water domain.** (a) Rise in temperature due to US in the skull and transcranial water domain at the end of the US stimulation waveform. The horizontal line at y = 5 mm marks the border of the skull. Note that heat is primarily generated in the skull adjacent to the US transducer. (b) Rise in temperature due to US in the transcranial water domain. Note that the temperature increase is much less than 1/100$^{th}$ of a degree. (c) Increase in temperature over time in the transcranial domain (upper plot) and intracranial domain (lower plot). Temperatures within the transcranial water domain are shown at the location of maximum transcranial intensity (black), and 1 and 2 mm lateral to that point (blue and red). The intracranial temperature is shown at the point of maximum intensity within the bone layer.

### 3.2. Model extension to homogenous brain tissue and curved skull surfaces

The transcranial water domain was then given material properties of brain tissue (white and gray matter of table 1) to compare how the profiles of intensity and temperature change based on the transcranial domain material. The intensity and heating of the cranial domain changed minimally as a result of this material change. As shown in figure 5, the general shape of both the intensity and heat profiles changes negligibly. Additionally, the maximum intensities and their locations between water and brain transcranial domains with a planar skull differ minimally (table 3). However, the maximal change in temperature in the transcranial domains differ by orders of magnitude, and their y-coordinates differ by a few millimeters as well, due to the large difference in attenuation coefficients between the two materials. The temperature increase due to the change of the transcranial domain from water to brain tissue was about 50 fold, similar to the 46 fold increase in attenuation coefficient in table 1. Additionally, changing the material properties of the transcranial domain slightly altered the magnitude of intensity effects in the cranial domain though it did not alter its heating behavior (table 3).

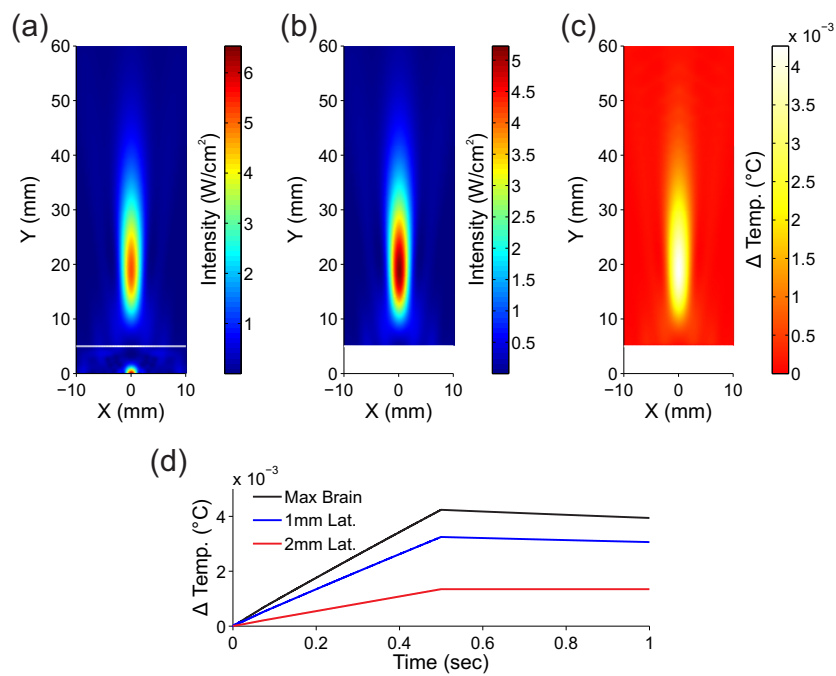

**Figure 5. Model of tFUS into a brain tissue domain.** (a) Intensity profile within both brain and skull layers. Note that peak intensities are within the skull layer, adjacent to the center of US transducer. (b) Intensity profile within the brain layer solely. (c) Heat generation by tFUS within the brain layer. Note that while heat generation is greater than in water, the temperature increase is still less than 1/100$^{th}$ of a degree. (d) Increase in temperature over time in the brain tissue domain at the location of maximum transcranial intensity (black), and 1 and 2 mm lateral to that (blue and red).

**Table 3.** Transcranial maxima and locations in computational models.

| Model | $I_{RMS}$ | | | Temperature Increase | | |
|---|---|---|---|---|---|---|
| | Skull Max (W/cm$^2$) | Trans Max (W/cm$^2$) | Trans Coordinate (mm) | Skull Max (°C) | Trans Max (°C) | Trans Coordinate (mm) |
| Plane Skull - Water | 6.81 | 5.22 | 19.74 | 0.15 | 8.2e-5 | 20.31 |
| Plane Skull - Brain | 6.53 | 5.22 | 19.21 | 0.14 | 4.3e-3 | 18.85 |
| Curved Skull - Brain | 1.64 | 3.20 | 18.68 | 0.03 | 2.4e-3 | 18.85 |
| Plane Skull - Layered Cortex | 6.51 | 5.25 | 19.00 | 0.14 | 4.3e-3 | 19.52 |

The skull layer was then changed to a circular arc to simulate a curved region of the skull and alter the coupling with the ultrasound transducer face in a manner that could similarly occur when placing the transducer on human skull. The resultant intensities and heat generation following curvature of the skull layer are shown in figure 6, where the range of intensities and temperature increases have been decreased in comparison to previous models with a planar skull layer. Additionally, the maximum intensity was no longer located in the skull layer, though the maximum temperature increase still occurred in the skull layer (table 3). The location of the maximum intensity in the brain domain moved down approximately 0.5 mm following curvature of the skull, though interestingly the location of maximum temperature increase did not change. Contours of the intensity profiles were analyzed and are shown in figure 7. The curvature of the skull results in a more compact region of high intensities at the ultrasound transducer's region of focus. The half maximum contour of intensity for the planar skull enclosed an area of 47.4 mm$^2$, while the curved skull decreased the enclosed area to 45.0 mm$^2$. Additionally, the planar skull produced a longer, thinner half max contour compared to the curved skull. Overall, the curved skull further focused the US beam to produce a more compact region of maximal effects, albeit at a lower magnitude compared to the planar skull.

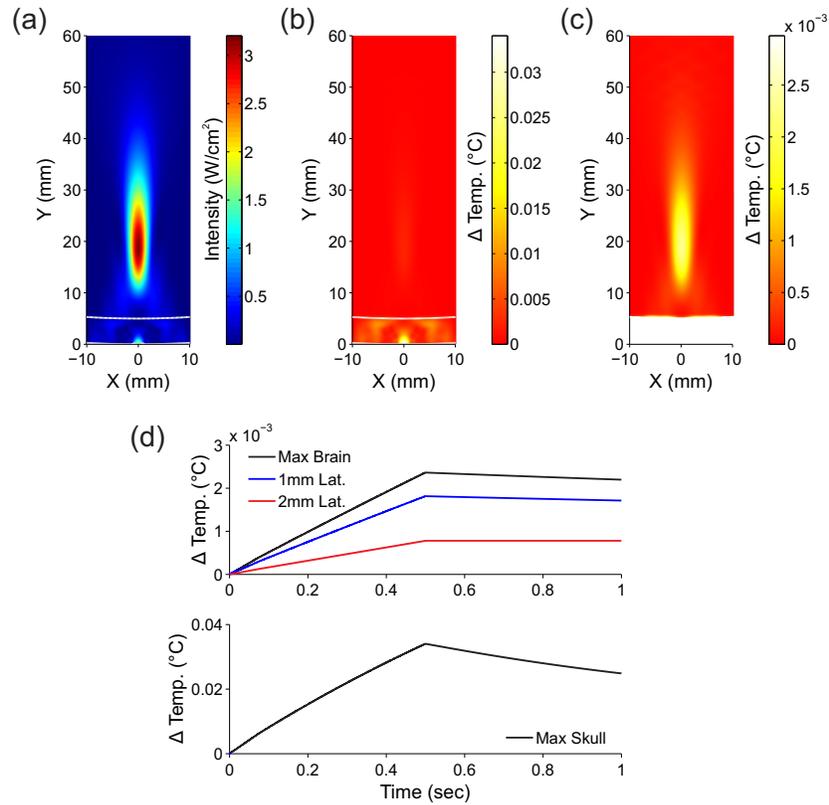

**Figure 6. Model of tFUS through a curved skull layer and into a brain tissue domain.** (a) Intensity profile within both the brain and skull layers. Note that the peak intensities are lower than for the model with a planar skull layer, and are no longer within the skull layer but are now within the focal region of the US transducer. (b) Profile of temperature increase within the brain and skull layers. Note that heating in the brain layer is still relatively low compared to that occurring in the skull. (c) Temperature increase within the brain domain. Note that the brain tissue at the interface with the skull is being heated at a similar level as that at the focal region of the US transducer. (d) Increase in temperature over time in the transcranial domain (upper plot) and intracranial domain (lower plot). Increases in temperature over time in the brain tissue domain are shown at the location of maximum transcranial intensity (black), and 1 and 2 mm lateral to that (blue and red). Note that there is less heat generation as compared to the model with a planar skull.

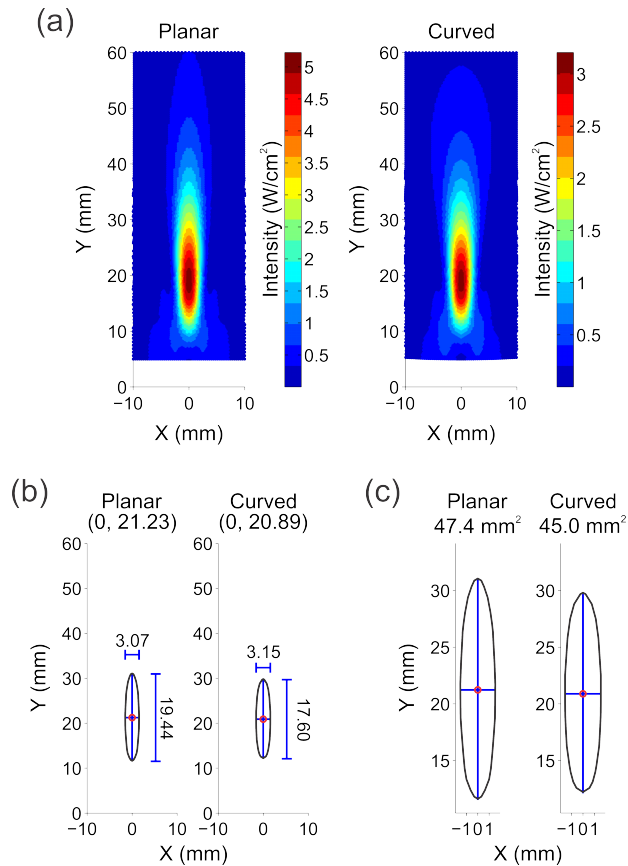

**Figure 7. Comparison of intensity profiles between planar (left) and curved (right) skull layers.** (a) Intensity contours of tFUS through a planar and curved skull layer. Note that the planar skull layer produced higher intensities and more elongated contours. (b) Half maximum intensity contours. Note that the planar skull layer produced a greater thresholded area with a longer major axis and an elevated centroid along the Y-axis. (c) Close up of half maximum intensity contours.

### 3.3. Model extension to layered cortical tissue and sensitivity analyses

Layers for CSF, gray matter, and white matter were added following a planar skull layer to further investigate the modulation of human cortex using focused ultrasound. The profiles of intensity and temperature rise are shown in figure 8, which did not change in overall shape compared to the previous homogenous brain model. The half maximum intensity contours are also very similar, with the homogenous brain model only having a slightly longer elliptical profile (by 0.19 mm) than the layered cortical model. Interestingly the maximum intensity in the skull layer decreased while the transcranial maximum increased in the layered cortical model compared to the homogenous brain (table 3). Between the cortical and homogenous models, the max temperature increases were very similar, with the only difference being that the cortical model's location for max temperature increase was elevated 0.67 mm along the Y-axis. Thus, the addition of a CSF layer between the skull and brain domains subtly influences the magnitude of intensities in both the skull and brain domains, but does not alter the magnitude of heat generation in either domain.

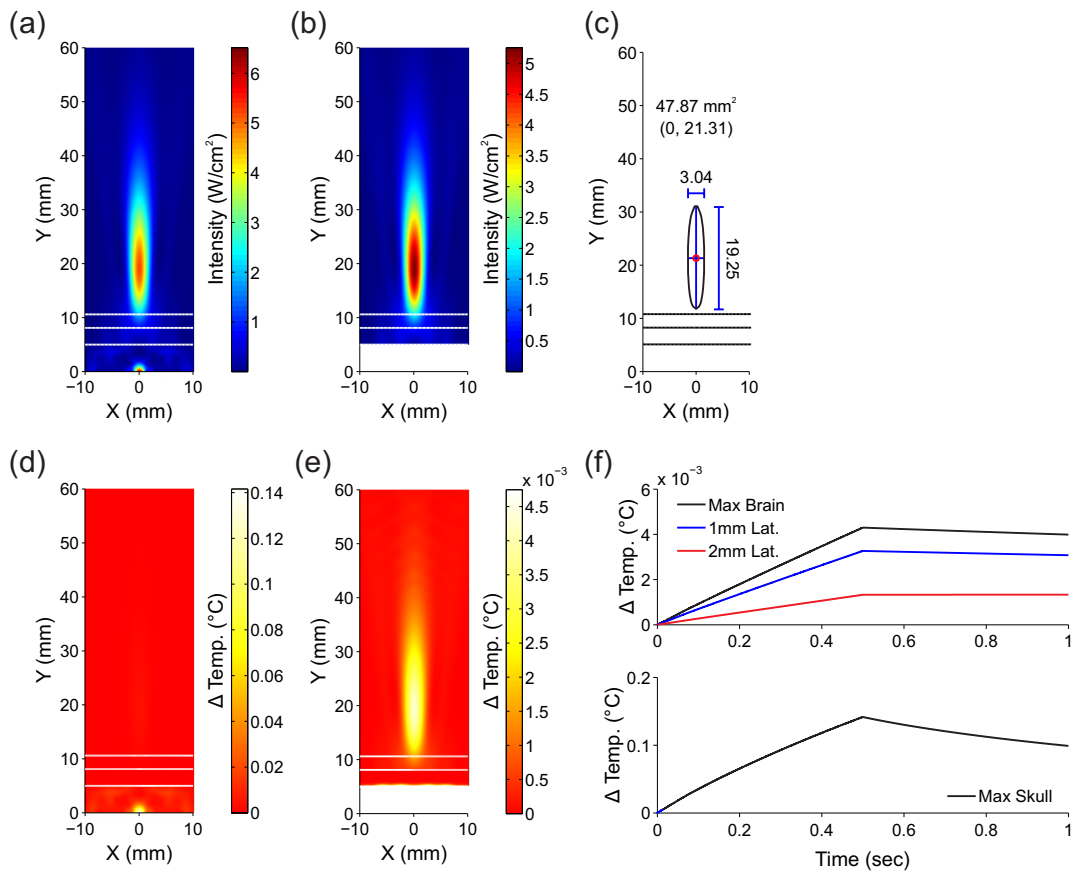

**Figure 8. Model of tFUS through a planar skull and layered cortex.** (a) Intensity profile within all layers with borders marked in white. Note that peak intensities are within the skull layer, adjacent to the center of US transducer. (b) Intensity profile within the transcranial layers solely. (c) Half maximum intensity contour. (d) Heat generation by tFUS within all layers. (e) Temperature increase in the transcranial layers from tFUS. (f) Increase in temperature over time in the transcranial domain (upper plot) and intracranial domain (lower plot). Temperatures within the cortex are shown at the location of maximum transcranial intensity (black), and 1 and 2 mm lateral to that point (blue and red). The intracranial temperature is shown at the point of maximum intensity within the skull layer.

As we found sparse literature on the differing acoustical properties of the cortical layers, sensitivity analyses were run to explore their impact on model behavior. The attenuation coefficient of white matter has been reported to be 1.4 times that of gray matter by Kremkau and colleagues [22], and after running simulations at a range of scaling factors we found that only for factors greater than 2.0 for the white matter attenuation coefficient did the half maximum intensity proxy for stimulation change noticeably. Overall, with increasing white matter attenuation coefficient, the elliptical profile decreases in area (figure 9a). The maximum intensity and temperature rise values in the skull and CSF did not change, but in the cortical tissue the maximum intensity decreased while the heat generation increased, albeit only slightly across the entire range (figure 9b). Interestingly, the elliptical profile of intensity did not change though the profile of temperature rise did (figures 9c-9e). The location of maximum temperature increase was located in the lower gray matter layer for the lowest two values of white matter attenuation coefficient simulated, after which the location of the maxima elevated into the white matter with a variation of about 1 mm as the attenuation coefficient continued to increase. The elliptical profile of temperature increase also returned for these higher values of attenuation coefficient. Overall, changes to the white matter attenuation coefficient between half and double that of the gray matter minimally changes the profiles of intensity and temperature change in the layered model of cortical tissue. Scaling of the attenuation coefficient of the gray matter (figure 10) had even less of an impact on the intensity profile of ultrasound compared to scaling white matter's coefficient (figure 10a), and produced a bimodal distribution of heat generation for high values of attenuation coefficient in the gray matter (figure 10e).

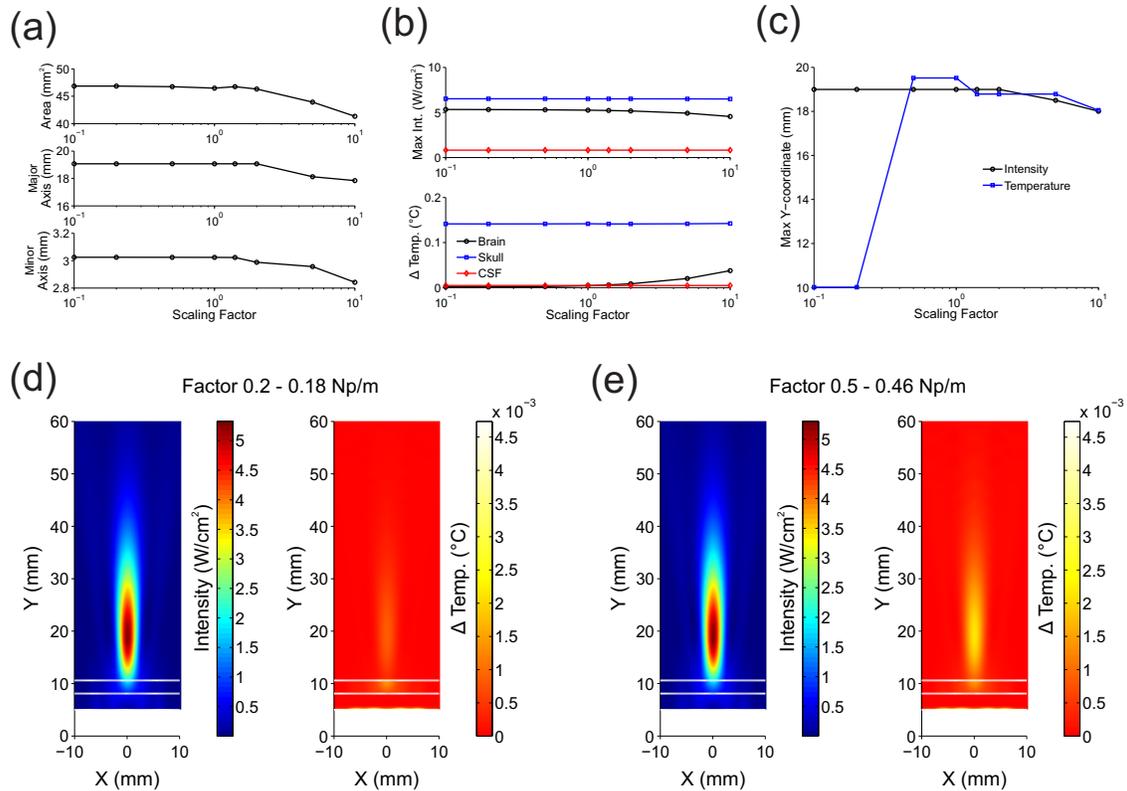

**Figure 9. Sensitivity analysis of the attenuation coefficient of white matter.** (a) Characterization of half maximum intensity contours. Overall, the area of the elliptical profile decreased with increasing the attenuation coefficient. (b) Max intensities and max temperature increases in cortical layers (black), skull (blue), and CSF (red). Measurements in the skull and CSF were insensitive, but as the attenuation coefficient of white matter increased, the max cortical intensity decreased while heat generation increased. (c) Location of the maximum intensity (black) and maximum temperature change (blue). Note that the jump in location of max temperature change between the scaling factors of 0.2 and 0.5 is due to a change in location from the gray matter to the white matter. (d) Intensity profile (left) and rise in temperature (right) for a white matter scaling factor of 0.2. Note that the maximum rise in temperature happens in the layer of gray matter. (e) Intensity profile (left) and rise in temperature (right) for a white matter scaling factor of 0.5.

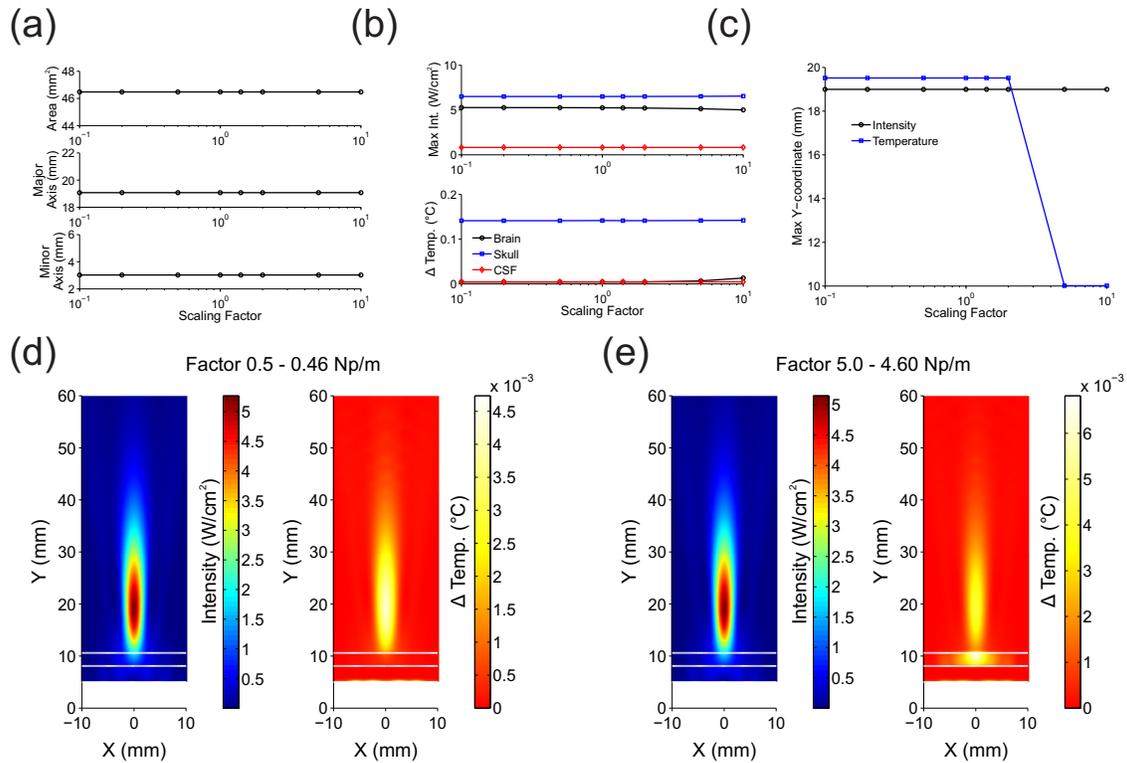

**Figure 10. Sensitivity analysis of the attenuation coefficient of gray matter.** (a) Characterization of half maximum intensity contours. Note that the intensity contour was insensitive to changes. (b) Max intensities and max temperature increases in cortical layers (black), skull (blue), and CSF (red). Metrics were largely insensitive. (c) Location of the maximum intensity (black) and maximum temperature change (blue). Note that the jump in location of max temperature change between the scaling factors of 2.0 and 5.0 is due to a change in location from the white matter to the gray matter. (d) Intensity profile (left) and rise in temperature (right) for a gray matter scaling factor of 0.5. (e) Intensity profile (left) and rise in temperature (right) for a white matter scaling factor of 5.0.

The initial inclusion of a CSF layer between the skull and brain domains subtly influenced the profiles of intensity and temperature increase in both the skull and brain domains. To further explore the effect of CSF layer presence, a sensitivity analysis of CSF layer thickness was run to explore the impact on model behavior. Regarding the area enclosed by the half maximum intensity contour (figure 11a), the thickness of the CSF layer minimally impacted the area enclosed, except for the case when the CSF was 15.5 mm thick. During this case of a very thick CSF layer, intensities within the half maximum intensity threshold were present within the CSF layer as well, reducing the area contained in brain tissue layers. The maxima intensity and temperature increases in the skull were most sensitive to the thickness of the CSF layer (figure 11b), with the skull domain having a greater range of values (0.82 W/cm$^2$ and 0.02°C) compared to the brain domain (0.40 W/cm$^2$ and 3.2e-4°C). The increase of maximal intensity within the CSF with scaling factor is attributed to more of the focal region of ultrasound being contained in CSF with increasing layer thickness. Additionally, the thickness of the CSF layer readily influenced the location of maxima, with maxima increasing in elevation as the CSF thickness increased (figures 9c-9e).

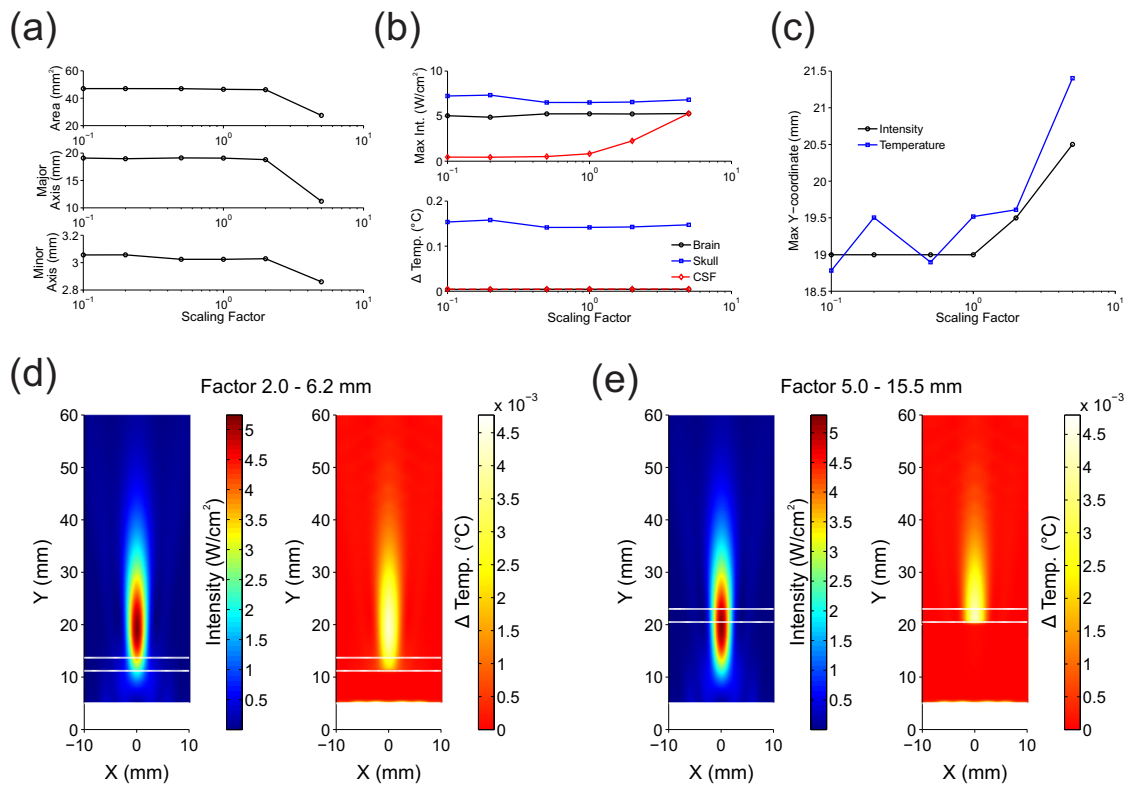

**Figure 11. Sensitivity analysis of the CSF layer thickness.** (a) Characterization of half maximum intensity contours. Overall, the area of the elliptical profile decreased with increasing the attenuation coefficient. Note though that at a scaling factor of 5.0, the half maximum intensity contour extends into the CSF layer. (b) Max intensities and max temperature increases in cortical layers (black), skull (blue), and CSF (red). Note that maximal values in the skull layer decreased slightly with increasing CSF layer thickness, while maximal values in the cortical layers were relatively less sensitive. (c) Location of the maximum intensity (black) and maximum temperature change (blue). Note that the increasing of location of maxima is namely due to the shift in CSF layer thickness. (d) Intensity profile (left) and rise in temperature (right) for a CSF thickness scaling factor of 2.0. (e) Intensity profile (left) and rise in temperature (right) for a CSF thickness scaling factor of 5.0.

Sensitivity analyses of the density and sound velocity properties of the gray and white matter layers revealed an interplay between all the model layers of skull, CSF, and gray/white matter influencing all measurements of interest (figure 12). The sound velocity of the white matter layer readily influences the ultrasound beam profile, as well as the maximal intensity in the skull (figure 12a). Varying the density of white matter caused similar trends in the change of maximal intensity and temperature values as the velocity, but influenced the half maximum intensity contour much less (figure 12b). The sound velocity of the gray matter layer also readily influences the ultrasound beam profile (figure 12c), with increasing half maximum contour area with increasing sound velocity, an opposite trend compared to the white matter sound velocity. Varying the density of gray matter (figure 12d) also caused a similarly oppositely sloped trend affecting the half maximum intensity contour compared to varying the density of white matter.

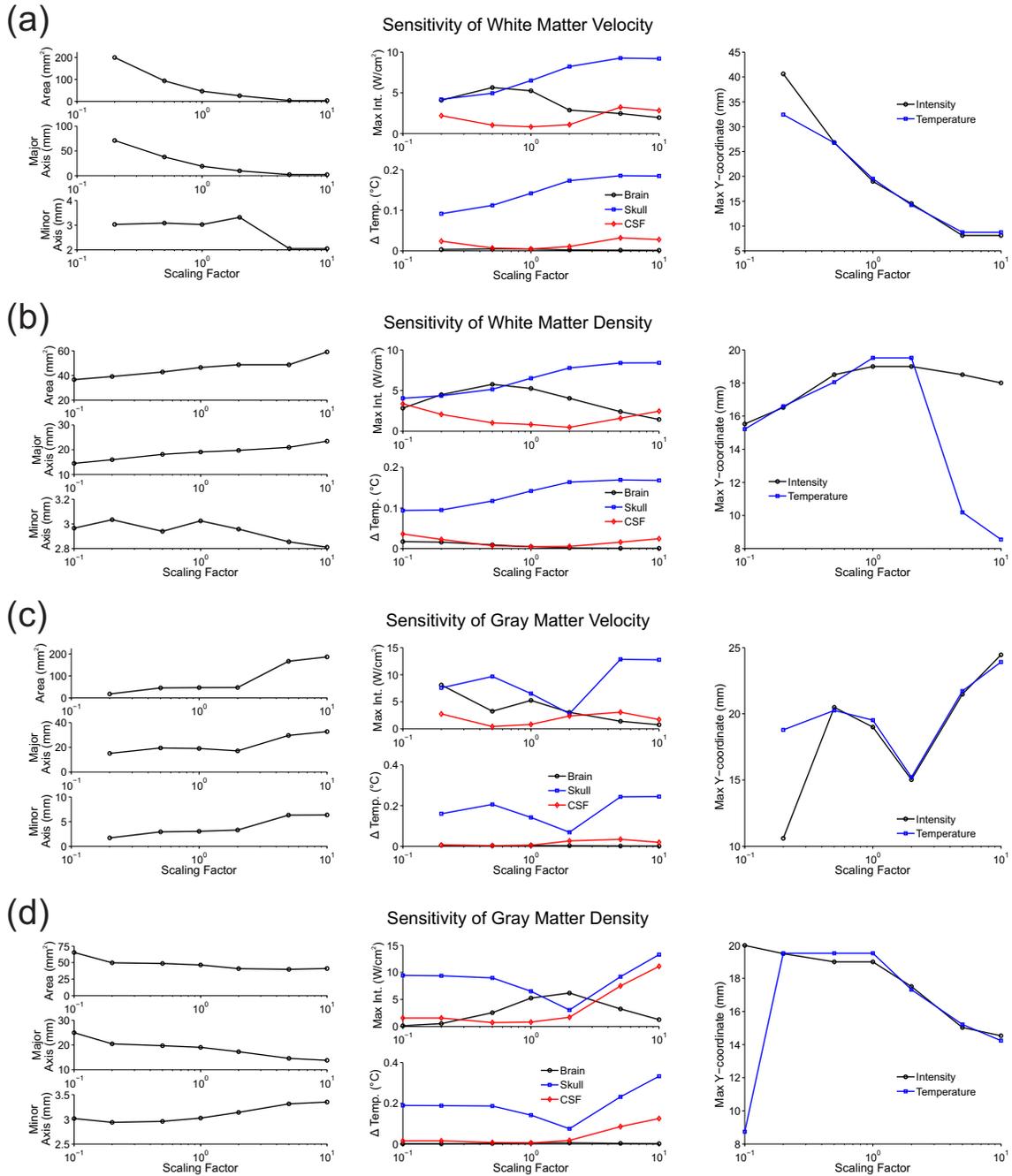

**Figure 12. Compilation of sensitivity analyses varying the density and speed of sound in the white and gray matter layers.** (a) Sensitivity of the half maximum intensity contour (left), maximal values (center), and y-coordinate of maximal values in the brain layers (right) in models varying the speed of sound in the white matter. (b) Sensitivity of measures in models varying the density of the white matter. (c) Sensitivity of measures in models varying the speed of sound in the gray matter. (d) Sensitivity of measures varying the density of the gray matter.

### 3.4. Influence of gyral geometry on the US beam

Transcranial transmission of the focused US beam results in an elliptical focus that can be deformed by the inclusion or proximity of sulci. These deformations are more qualitatively apparent in the models of the slanted sulci, as at certain horizontal offsets the beam focus spans across the CSF of sulci into the brain regions on either side (figure 13b). The quantification of the half maximum intensity contours is shown in figure 14 to allow comparison of the deformations to the US beam by gyral anatomies, as well as to a baseline comparison point derived from a model with no sulci ('NS' in figures). Overall, the more the beam focus is localized to a sulcus, the smaller the thresholded area and axes lengths compared to other cases in models including sulci. This is especially notable in the model with perpendicular sulci, where a horizontal offset of 6 mm, aligning the transducer well with a sulcus, results in a drop of area of about 40 mm$^2$ compared to the rest of the offsets (figure 14a).

    Particularly noteworthy though, is that the absence of sulci in the model can result in a decreased area of stimulation compared to models with sulci. This is shown well in the comparison of the US beam focus in the model with no sulci to the beam focus in the model with perpendicular sulci. Except for the case when the region of focus is centered on a sulci itself, when the region of focus is either between two sulci or near one it results in greater areas of stimulation compared to the model with no sulci (figure 14a). Also particularly noteworthy is that the gyral anatomy did not typically result in sufficient deformation of the US beam focus to rotate the centroidal principal moment axes by more than one degree. Only one particular offset with rotated sulci (9 mm) managed to rotate the principal moment axis by more than one degree (figure 14c).

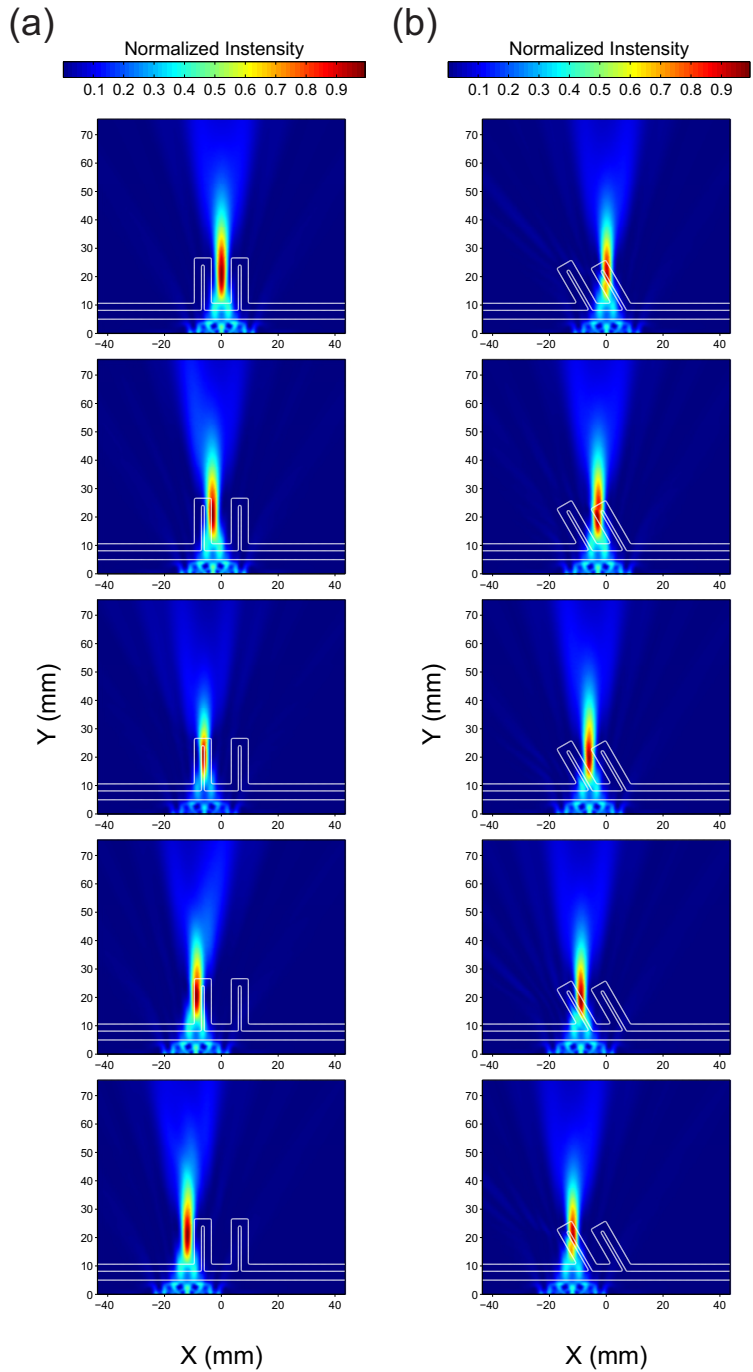

**Figure 13. Modeling focused ultrasound in simplified gyral anatomies.** Normalized intensity profiles for perpendicular (a) and slanted (b) sulci, with 0, 3, 6, 9, and 12 mm transducer offsets from the x-origin (top to bottom respectively). Note that the transitions between CSF and neighboring tissue result in deformations of the focused US beam as compared to the previous planar, layered models.

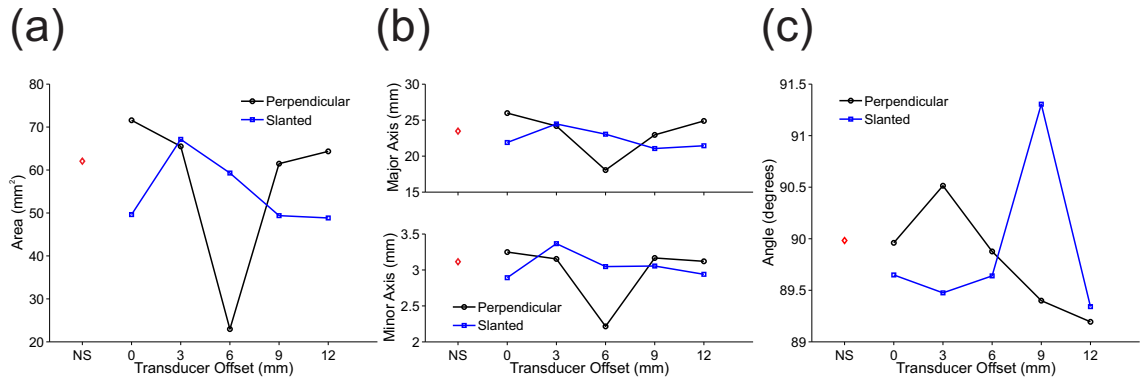

**Figure 14. Characterization of gyral geometry effects on the focused ultrasound beam.** Following thresholding of the intensity profiles, the resultant elliptical profiles were characterized using area (a), axis length (b), and centroidal principal moment axis angle (c) to quantify the effect of perpendicular (black line) and slanted (blue line) gyral anatomy with varying transducer offsets on the focused US beam. These parameters are also compared to a similar layered cortical model with no sulci (labeled 'NS' along x-axis, red diamond in graphs) to establish a baseline free of gyral effects for comparison. Overall, when the ultrasound beam was focused in a sulcus, the thresholded area and axes lengths decreased. However the angle of the major axis was minimally perturbed by gyral anatomy.

## 4. Discussion

Transcranial focused ultrasound is an appealing approach for noninvasive neuromodulation of cortical tissue for a wide variety of applications, including those with deeper cortical targets. However, the improvement and adoption of ultrasound methods for neurostimulation is greatly dependent on furthering our understanding of ultrasonic mechanisms, including its insertion behavior across the skull. We developed a computational model of the resultant intensity profiles of transcranially focused ultrasound based on acoustic tests in a water tank, and extended the model to solve for the heating by the ultrasonic neuromodulation waveform. We used the model to then explore the effect of tissue properties and model geometries on the behavior of the ultrasound beam. To quantify the model response of ultrasound, we characterized the ultrasound beam using half maximum intensity contours and their corresponding area moments of inertia. While the relationship between ultrasound intensity and stimulation of neural tissue is not established, the half maximum intensity contours provided a quantitative measure of the model response to estimate the influence of tissues and geometry on the region of effects by tFUS. By beginning to investigate and consider these issues in the context of neuromodulation, we can advance the utility of focused ultrasound methods for human neuromodulation.

To ensure that the computational models would be relatively accurate and credible, we began with construction of a computational model recreating acoustic testing of tFUS in a water tank. While it is possible to adjust the computational model to have an identical maximum intensity value as that observed in the experimental measurements, obtaining an identical profile of intensities is more difficult. Most notable in the difference between the computational and experimental profiles of intensity in figure 2 is the warped region of moderately high intensities near the inner surface of the skull and below the maxima. This may be largely attributed to differences between the experimental approach and the idealized computational model, namely the inhomogeneous, anisotropic, and slightly curved human skull fragment used in the experimental tests. Unlike the idealized skull layer in the computational model, human skull has an inhomogeneous curved structure with a varying density and thickness that is compensated for in applications requiring very precise control of the transcranial distribution of ultrasound [27, 28]. Outside of the region near the inner skull surface, at the focal region and far field locations, the intensity profiles between the experimental and computational models are qualitatively similar and we deemed the computational model an acceptable recreation of the experimental observations for the purposes of this investigation.

Using the computational model we were also able to calculate the intensities within the skull layer, and simulate the heat generation from focused ultrasound in both the cranial and transcranial domains. The majority of heating takes place in the skull layer, largely due to the fact that the attenuation coefficient of the skull is much higher than that of the water. In fact, the model overestimates the heating of the transcranial water domain, as we used a value of 0.02 Np/m, while the attenuation coefficient of water at room temperature is closer to 6e-3 Np/m based on reported data [29]. We used this larger value due to our representing CSF with the same parameter set in later models, as the density and sound velocity of water were found to be similar to that of CSF according to one source [30], and the CSF containing proteins and other compounds likely increases the attenuation coefficient to some degree.

Transcranial magnetic stimulation is another form of noninvasive neuromodulation that passes unimpeded through skull and whose manifestation of effects (electric fields) is influenced by the geometry of neural tissue [2]. As reflected in the simulations of this work though, tFUS seems to be manipulated in an opposite manner compared to TMS; the skull is the barrier to transmission of energy by ultrasound and the geometry of neural tissue only influences the manifestation of effects due to ultrasound (i.e. intensity and heating) subtly. As the properties of the skull (e.g. thickness, density, curvature) can vary over the expanse of the cranium [28], this implies that the transcranial effects of US can vary with transducer placement on the skull. Indeed, curvature of the skull layer in the computational model resulted in a 62% drop in maximal transcranial intensity, and a 56% drop in transcranial heat generation. The drop in effects by ultrasound within the skull layer was of an even greater scale, though they are not of concern in regards to neuromodulation but merely as a safety check and possible means of secondary effects. The influence of tissue geometry on the effects of US were quantified using the half maximum intensity contours, and was found to have subtle effects. Overall, the region of effects by US stayed at the focus of the ultrasound transducer, with CSF in sulci being the source of subtle influence on the geometry of the intensity contours. This translates into the targeting of US for neurostimulation not being variable with the intracranial geometry and thus not being a significant concern for the design of an ultrasound transducer's region of effects. In comparison to TMS where the induced electric fields are on the scale of centimeters and greatest at the gyral crown, but the effective electric field for neuronal stimulation is for elements within the gyral walls [31], one can focus design efforts of an ultrasound transducer to have a region of focus at the desired depth after accounting for the placement of the transducer on the cranium. Additionally, differences in tissue properties between white and gray matter could have a more substantial influence on the ultrasound beam, if the properties between the two differ greatly due to developmental or pathological changes. Thus placement and targeting of the ultrasound transducer are the primary factors of concern when applying tFUS, especially as the scale of effects by tFUS are in the range of millimeters.

Overall, the intracranial manifestation of effects by US (intensity, heating) is more readily controlled than the effects of TMS and other electromagnetically based noninvasive neuromodulation methods. The profile of these manifestations though depends on the insertion behavior of US across the various layers of tissue. As US offers the advantages of finer spatial resolution and variable depths of stimulation compared to noninvasive electromagnetic methods though, it is an appealing alternative to electromagnetic methods for a number of possible applications. However, it is still not clear how the neuronal response couples to the exertions of ultrasound on tissue. Using a computational model to systematically investigate parameters of interest, we found that the profiles of intensity produced by tFUS is relatively insensitive to the geometry of intracranial tissue, that the material properties of the intracranial tissue can influence the intensity profile more substantially, and that the skull is a major source of influence on the ultrasound beam profile. An understanding of both the insertion behavior of ultrasound across the skull and the coupled neuronal response is key to the continued advancement of ultrasound stimulation methods.

**Disclosure**